\documentclass[12pt,a4paper,dvips,twoside]{article}
\usepackage{a4p}                 % use mine - marginpar parameters slightly different
\usepackage{cite,mcite}
\usepackage{graphicx}
\usepackage{physics }
\usepackage{amsmath}
\usepackage{amssymb}
\usepackage{pennames}
\usepackage{l3_title}
\usepackage{ifthen}
\date{October 4, 2001}
\preprint{2001-072}
\newlength{\capindent}
\newlength{\capwidth}
\newlength{\figwidth}
\newcommand{\icaption}[2][!*!,!]{\hspace*{\capindent}%
  \begin{minipage}{\capwidth}
    \ifthenelse{\equal{#1}{!*!,!}}%
      {\caption{#2}}%
      {\caption[#1]{#2}}
  \end{minipage}}
\newcommand{\Pgtp}{\ensuremath{\Pgt^+}}%    tau+ not in pennames
\newcommand{\Pgtm}{\ensuremath{\Pgt^-}}%    tau- not in pennames
\newcommand{\PZ}{\ensuremath{\mathrm{Z}}}%      Z not in pennames
\renewcommand{\Pqu}{\ensuremath{\mathrm{u}}}%   u replace q_u in pennames
\renewcommand{\Pqd}{\ensuremath{\mathrm{d}}}%   d replace q_d in pennames
\renewcommand{\Pqc}{\ensuremath{\mathrm{c}}}%   c replace q_c in pennames
\renewcommand{\Pqs}{\ensuremath{\mathrm{s}}}%   s replace q_s in pennames
\renewcommand{\Pqb}{\ensuremath{\mathrm{b}}}%   b replace q_b in pennames
\newcommand{\JETSET}{{\scshape jetset}}
\newcommand{\ARIADNE}{{\scshape ariadne}}
\newcommand{\HERWIG}{{\scshape herwig}}
\newcommand{\PYTHIA}{{\scshape pythia}}

\newcommand{\LEP}{{\scshape lep}}

\newcommand{\SLD}{{\scshape sld}}
\newcommand{\Lthree}{{\scshape l}{\footnotesize 3}}

\newcommand{\Nch}{{\ensuremath{n}}}                        % Nch
\newcommand{\Hq}{{\ensuremath{H_q}}}                       % H_q
\baselinestretch        % allow stretching of space between lines as well as between paragraphs
 
\begin{document}
\begin{titlepage}
\title{Measurement of      \\ charged-particle multiplicity
         distributions \\
         and their \boldmath{$H_q$} moments
         in \\
        hadronic Z decays \\  at LEP}
 
\author{The L3 Collaboration}
%
% The abstract
%
\abstract{The charged-particle multiplicity distribution
    is measured  for all hadronic events as well as for light-quark and \Pqb-quark events
    produced in $\Pep\Pem$ collisions at the \PZ\ pole.
    Moments of the charged-particle multiplicity distributions are calculated.
    The \Hq\ moments of the multiplicity distributions are studied, and their
    quasi-oscillations as a function of the rank of the moment are investigated.
}
 
\submitted        %  adds "Submitted to ..."

\end{titlepage}
 
\newpage
%{\pagestyle{empty}
% \cleardoublepage
%}
\setcounter{page}{1}

\section*{Introduction}
 
Since quarks and gluons are not observed directly,
the understanding of the hadronization process
whereby a quark-gluon system evolves to hadrons is of importance
and provides a tool for studying the quark-gluon system itself.
One of the most basic characteristics of the resulting hadronic system
is the distribution of the number of hadrons produced.
 
Assuming local parton-hadron duality (LPHD)\cite{lphd},
characteristics of the charged-particle multiplicity distribution
are directly related to the characteristics of the corresponding parton distributions.
The parton distributions are calculable using perturbative quantum chromo-dynamics (pQCD).
In particular, the dependence  on the center-of-mass energy, $\sqrt{s}$,
of the mean, $\langle n\rangle$,
of the charged-particle  multiplicity
is an important test of pQCD.
Since these calculations are only valid for light quarks,
a separate measurement for light quarks is of interest.
 
In this Letter, the charged-particle multiplicity distributions
of hadronic decays of the \PZ\ boson are measured
for \Pqb- and for light-quark (\Pqu, \Pqd, \Pqs\ and \Pqc) events as well as for all events.
From these distributions moments are calculated, which characterize the shape of the distributions.

The shape of the charged-particle multiplicity distribution is
a fundamental tool in the study of particle production. Independent emission
of single particles leads to a Poissonian multiplicity distribution.
Deviations from this shape, therefore, reveal correlations~\cite{kittel}.
To study the shape, we use the normalized factorial moments.
In terms of the multiplicity distribution, $P(n)$, the  normalized factorial moment
of rank $q$ is defined by
\begin{equation}
   F_q = \frac{\sum_{n=q}^\infty n(n-1)...(n-q+1) P(n)}{\left(\sum_{n=1}^\infty n P(n)\right)^q} \quad.
\end{equation}
%The factorial moment of rank $q$ corresponds to an integral over the $q$-particle density and
It reflects correlations in the production of up to $q$ particles.
If the particle distribution is Poissonian, all $F_q$ are equal to unity.
If the particles are correlated,
the distribution is broader and the $F_q$ are greater than unity.
If the particles are anti-correlated,
the distribution is narrower and the $F_q$ are less than unity.
 
Normalized factorial cumulants, $K_q$, obtained from the normalized factorial moments by
\begin{equation}
  K_q = F_q - \sum_{m=1}^{q-1} \frac{(q-1)!}{m!\,(q-m-1)!}\,K_{q-m}F_m \quad,
\end{equation}
measure the genuine correlations between $q$ particles, \ie, $q$-particle correlations which are not
a consequence of correlations among fewer than $q$ particles.
 
Since $|K_q|$ and $F_q$ both increase rapidly with $q$, it is useful to define the $H_q$ moments,
\begin{equation}
   H_q = \frac{K_q}{F_q} \quad,
\end{equation}
which have the same order of magnitude over a large range of $q$.

\begin{figure}%[htbp]
  \begin{center}
   \includegraphics[width=.8\figwidth]{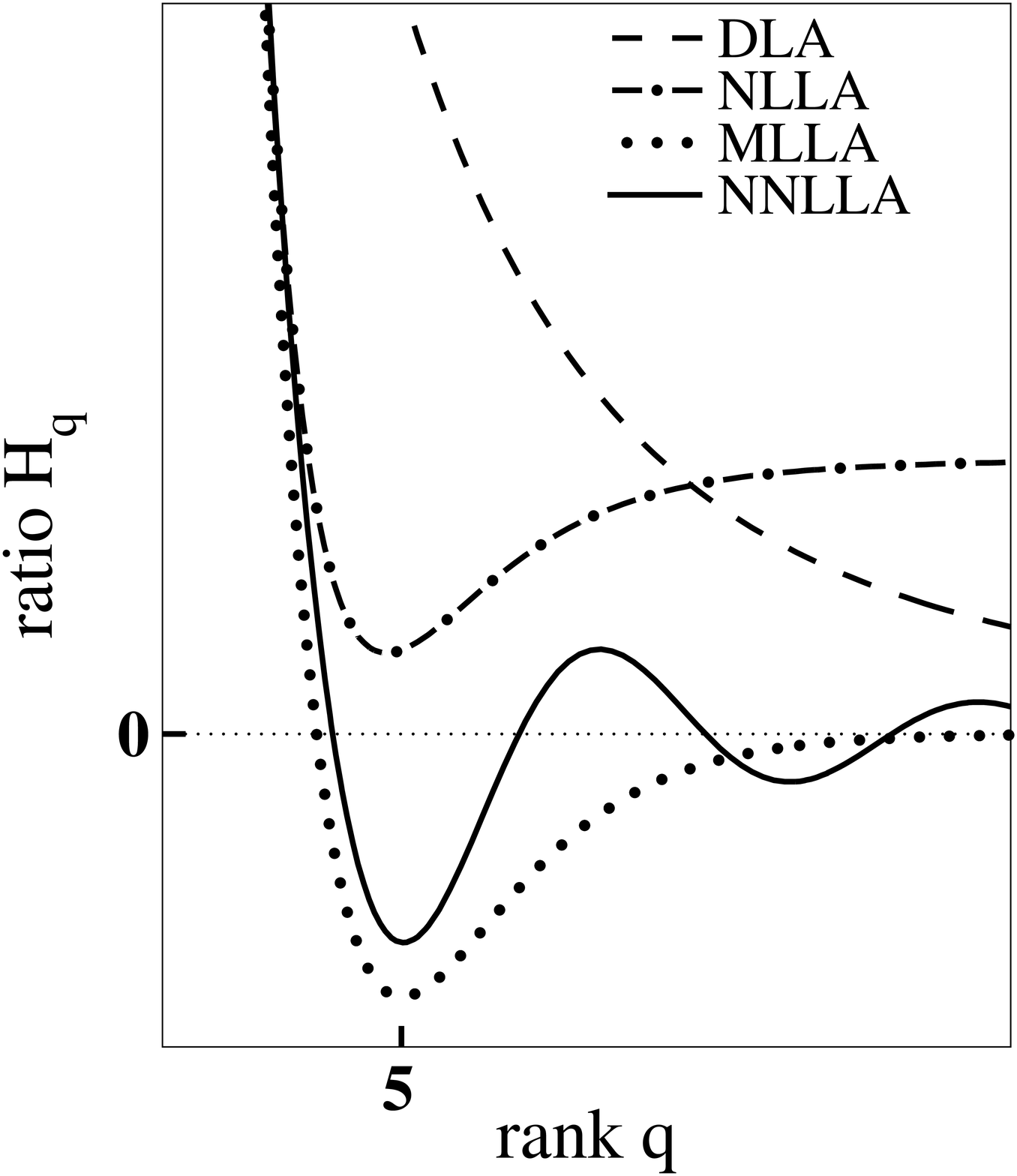}
  \end{center}
\icaption{Qualitative behavior of $H_q$ as a function of $q$ for various approximations of perturbative QCD
            [3,8].
%         \cite{dremin1,*dremin1a,dremin2}.   %   cite does not work within icaption !!!!!!!!!!!!!!!!
\label{fig:Hqpred}
}
\end{figure}

The shape of the charged-particle multiplicity distribution analyzed in terms of
the $H_q$  was found to reveal
quasi-oscillations~\cite{dremin2,dremin2a,SLD,shoulder},
when plotted versus the rank $q$, in \Pep\Pem, as well as hadron-hadron, hadron-ion and ion-ion
interactions.
In \Pep\Pem\ annihilation, this result was      interpreted
\cite{SLD,dremin_gary} in terms of pQCD, from which the $H_q$ of the parton multiplicity
distribution were calculated~\cite{dremin2,dremin1}.
The expected behavior  of $H_q$ \vs\ $q$ is quite sensitive to the approximation used,
as is illustrated qualitatively in Figure~\ref{fig:Hqpred} for
                                         the double           logarithm approximation (DLA),
                                         the modified leading logarithm approximation (MLLA),
                                         the  next-to-leading logarithm approximation (NLLA),
and                                      the next-to-next-to-leading logarithm approximation (NNLLA).
In the NNLLA  a negative first minimum is
expected near $q=5$ and quasi-oscillations about zero are expected for larger values of~$q$.
 
According to the LPHD hypothesis,
hadronization does not distort the shape of the  multiplicity distribution.
If this is valid, the same shape
may be expected for the charged-particle multiplicity distribution as for the
parton multiplicity distribution.

\section*{Experimental procedures}
\subsection*{Event selection}
This analysis is based on  1.5 million hadronic events
collected by the \Lthree\  detector\cite{det1} at \LEP\ in the years 1994 and 1995 at the \PZ\ pole.
 
Events are selected in a two-step procedure\cite{mangeol}.
First,
at least 15 calorimetric clusters of at least $100\MeV$ are required
in order to reduce background from the $\Pep\Pem\rightarrow\Pgtp\Pgtm$ process.
Hadronic events from the process $\Pep\Pem\rightarrow\Pq\Paq$
are then selected by requiring small energy imbalance both
along and transverse to the beam direction.
 
The second step is the selection of charged tracks
measured in the central tracker and the silicon micro-vertex detector.
A number of quality cuts are used to select well-measured tracks.
Further, the thrust direction calculated from the charged tracks is required to lie within the full
acceptance of the central tracker.
No selection specifically rejects or selects tracks from long-lived neutral particles.
The track selection efficiency, determined from Monte Carlo, is about 75\%.
The resulting data sample corresponds to approximately one million selected hadronic events,
%and has a purity greater than 99\%.
and has a purity of about     99.8\%.
 
To correct for detector acceptances and inefficiencies,
we make use of the \JETSET\ {\footnotesize7.4}\cite{JETSET74} parton shower Monte Carlo program,
tuned using \Lthree\ data. % \cite{tune,l3-38}.
Events are generated, passed through the \Lthree\ detector simulation
program\cite{sil3}, and further subjected to time-dependent detector effects.
Then they are reconstructed and the events and tracks are selected in the same
way as the data.
For systematic studies we also use events generated by \ARIADNE\ {\footnotesize4.2}\cite{ARIADNE}.
For comparisons with the data we use \HERWIG\ {\footnotesize5.9}\cite{HERWIG56} as well as \JETSET.
 
To select \Pqb- and udsc-quark enhanced samples, we use the full
three-dimensional information on tracks from the central tracker
to calculate for each track the probability that it
originated at the primary vertex \cite{l3-127}.
We select
\Pqb- and udsc-quark samples with
        purities of about 96\%  and 93\%
and efficiencies of about 38\%  and 96\%, respectively.
 
\subsection*{Unfolding}
The resulting multiplicity
distributions are fully corrected for detector resolution using an iterative Bayesian
unfolding method \cite{dagostini}.
The detector and generator level Monte Carlo events are used to construct
a matrix $R(n_\mathrm{det},n)$ which represents
the probability that $n_\mathrm{det}$ tracks would be detected if $n$ charged particles were produced.
A distribution, $P_0(n)$, is assumed for $n$.  For this $P_0$, the distribution
expected in the detector is $P_0^\mathrm{det}(n_\mathrm{det}) = \sum_{n} R(n_\mathrm{det},n) P_0(n)$.
This is compared to the actual distribution of the raw data,
and, making use of Bayes' theorem, % \cite{bayes},
an improved multiplicity distribution is calculated, which replaces $P_0(n)$ in the above expression.
This process is repeated iteratively until satisfactory agreement
between the expected and actual raw data distribution is found.
In practice, this occurs after the second iteration if the \JETSET\ multiplicity distribution
is chosen as $P_0(n)$.
 
In addition, corrections are made for efficiency and acceptance of the event selection,
initial state radiation, and \PKzS\ and \PgL\ decays.
Furthermore, the distributions for the \Pqb- and udsc-enhanced samples are corrected for the
purity of the flavor selection.
 
The unfolding method gives \cite{dagostini} an estimate of the covariance matrix of the
unfolded distribution.
This matrix, combined with the uncertainties on the corrections
mentioned above, is used to determine the uncertainties on the
moments of the multiplicity distribution for the all- and udsc-flavor cases.
When the statistics is too small, as in the b-flavor case, the  uncertainty on the
estimate of the covariance matrix is large.  In this case we use a Monte Carlo
method. Many Monte Carlo variations of the raw data multiplicity distribution are made,
choosing the number of events at each multiplicity from a Poisson distribution having
as mean the observed number of events.  These Monte Carlo distributions are then
analysed in the same way as the data distribution.  The uncertainty on a moment
is determined from the spread in values of the moments of the Monte Carlo distributions.
For the high-statistics cases, both methods agree.

\subsection*{Systematic uncertainties}
 
The following sources of systematic uncertainty are investigated:
 
\begin{description}
 \item{\bfseries Selection.}  The value of each cut used in the event selection is
     varied independently over a reasonable range around the default value
     and the resulting fully corrected distributions, together with their covariance matrices, determined,
     and from them the moments of the multiplicity distribution.
     For each multiplicity, as well as for each multiplicity moment, we assign
     a systematic uncertainty of half of the maximum difference between the new values.
     The same procedure is followed for the track selection and flavor tagging.
     For flavor tagging there is an additional contribution due to an uncertainty of 2.5\%
     in the purity of the resulting sample,
     which accounts for the different response of the tagging algorithm to data and Monte Carlo.
 \item{\bfseries Monte Carlo uncertainties.}  The analysis is repeated using \ARIADNE\ instead of \JETSET\
     to determine the corrections and the unfolding matrix.
     The difference between the two results is taken as the systematic uncertainty.
     Further, the \Pqc- and \Pqb-quark fragmentation parameters, $\epsilon_\Pqc$ and $\epsilon_\Pqb$,
     are varied.
     Also, the strangeness suppression parameter is varied by an amount consistent with the measured \PKzS\
     production rate\cite{PDG2002}.
     In each case, half the difference between the results using the two parameter values
     is taken as the systematic uncertainty.
 \item{\bfseries Unfolding method.} Three contributions are determined:
     First, \ARIADNE\ is used to derive the initial distribution.
     Secondly, the analysis is repeated using  a different number of iterations  in the unfolding.
     Finally,  the detector level multiplicity distribution of events generated by \ARIADNE\ is unfolded
     using the response matrix, $R(n_\mathrm{det},n)$, determined using \JETSET\ events.
     In each case, the difference from the default value  is taken as the systematic uncertainty.
 \item{\bfseries Background.} The background of about 0.2\% is mostly from two-photon processes.
     We take as a systematic uncertainty the effect of twice the amount of estimated background.
\end{description}
The contributions from each of these  sources are added in quadrature.
The track selection contributes the dominant part of the total systematic uncertainty
when all events are used, while the flavor-tagging purity uncertainty dominates that of the udsc sample.
For the b-quark sample, these two contributions are about equal.
 
In addition, the accuracy of the simulation of the rate of photon conversion is considered.
This is found to be about 15\% smaller than in data\cite{mangeol} and is assigned as a systematic
uncertainty on $\langle\Nch\rangle$.
It is found to be negligible for the other moments.
Breakdowns of the systematic uncertainties on $\langle\Nch\rangle$
are shown in Table~\ref{tab:syst}.
 
\begin{table}%[htbp]
\begin{center}
\begin{tabular}{|l|r@{.}l|r@{.}l|r@{.}l|}\hline
  Source           & \multicolumn{2}{c|}{full}
                   & \multicolumn{2}{c|}{udsc}
                   & \multicolumn{2}{c|}{\Pqb}  \\ \hline
  Event selection  &   0&005    &  0&006    &  0&004  \\
  Track selection  &   0&090    &  0&080    &  0&116  \\
  Tagging cuts     &\multicolumn{2}{c|}{\ } &  0&018 & 0&021 \\
  Tagging purity   &\multicolumn{2}{c|}{\ } &  0&185 & 0&126 \\
  MC modeling      &   0&032    &  0&031    &  0&040   \\
  Unfolding        &   0&034    &  0&034    &  0&043   \\
  Background       &   0&024    &  0&024    &  0&023   \\
  \Pgg\ conversion &   0&039    &  0&039    &  0&039   \\ \hline
\hline
  Total            &   0&11     &  0&21     &  0&19  \\
\hline
\end{tabular}\end{center}
\caption{Contribution of the various sources of systematic uncertainty to the measurement
  of the mean charged-particle multiplicity, $\langle\Nch\rangle$.
\label{tab:syst}
}
\end{table}
 
\section*{Results}
 
\subsection*{Charged-particle multiplicity distributions}
 
\begin{figure}[htbp]
  \begin{center}
    \includegraphics[width=\figwidth]{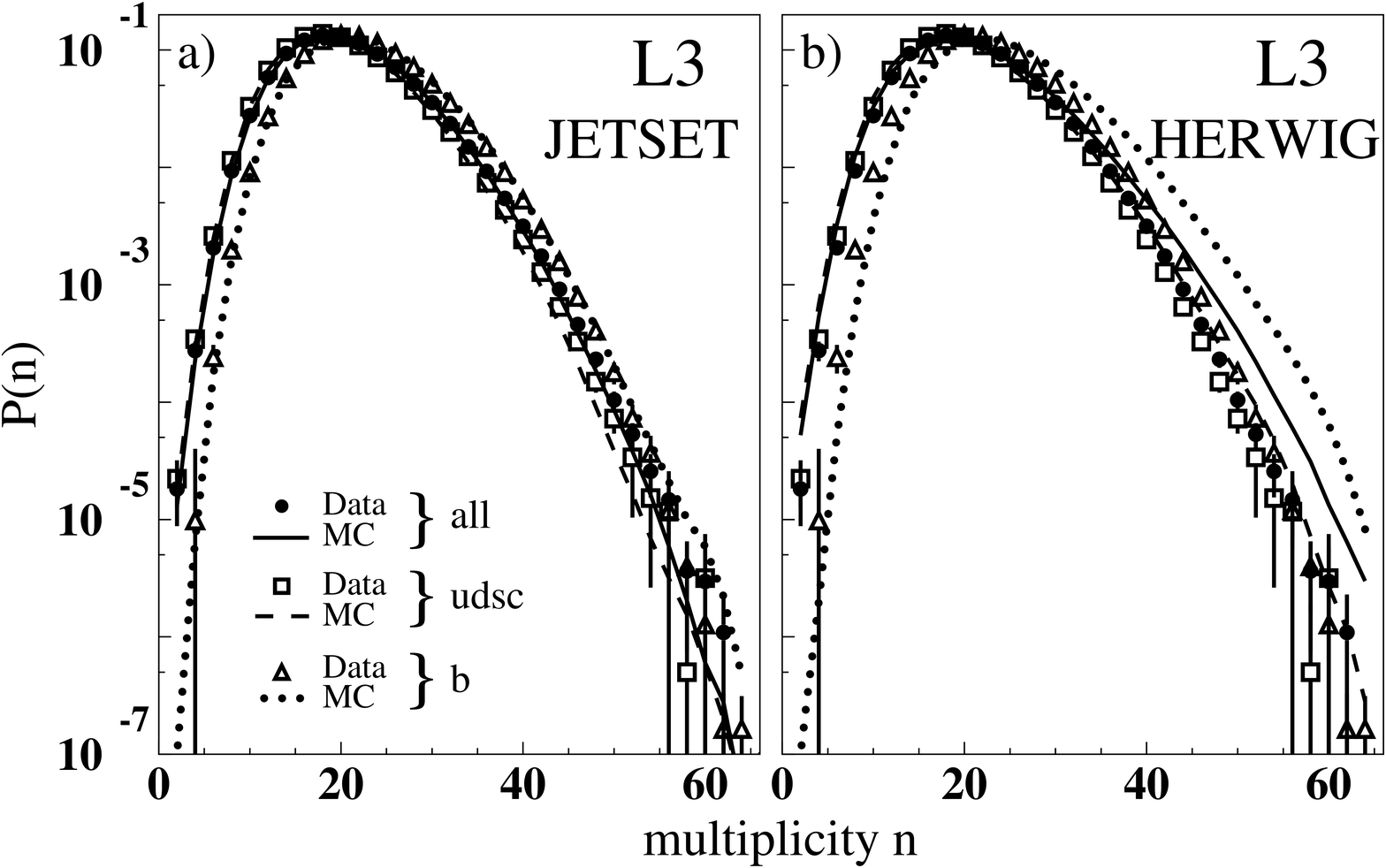}
    \includegraphics[width=\figwidth]{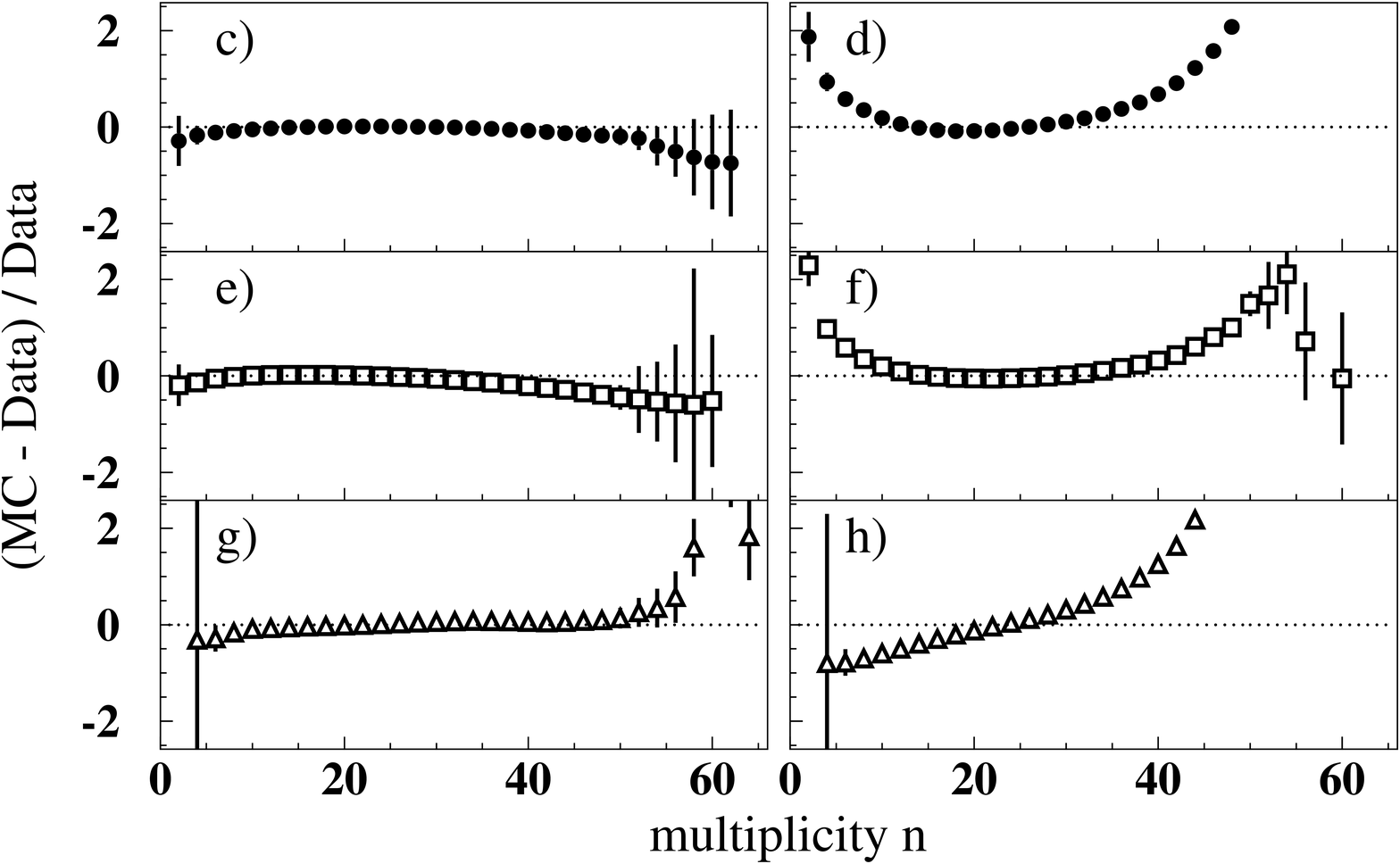}
  \end{center}
\icaption{
   \label{fig:mult1}
    Charged-particle multiplicity distribution for
    all, udsc-, and \Pqb-quark events
    compared to the expectations of (a,c,e,g) \JETSET\ and (b,d,f,h) \HERWIG.
  The error bars include both statistical and systematic uncertainties.
}
\end{figure}

Charged-particle multiplicity distributions are measured both including and excluding \PKzS\ and \PgL\ decay
products.\footnote{Note that \PgSm, \PgXm and \PgOm\ have only one charged particle among their decay
products apart from those produced in \PgL\ decay, and \PgSz\ and \PgXz\ have none. Thus including or not
the decay products of these baryons does not affect the charged multiplicity except through the \PgL\
decay.}
Figure~\ref{fig:mult1} shows the charged-particle
multiplicity distribution including \PKzS\ and
\PgL\ decay products for the full, udsc- and \Pqb-quark samples.
All distributions agree rather well with \JETSET, %\ {\footnotesize7.4} \scshape ps},
but in all cases \HERWIG\
gives a poor description of the data, as is seen in Figures~\ref{fig:mult1}a and~\ref{fig:mult1}b.
 
From these distributions
various moments of the charged-particle multiplicity distribution are calculated.
The results are summarized in Table~\ref{tab:t1t3}.
The mean multiplicity including \PKzS\ and \PgL\ decay products is consistent with our previous
measurements\cite{l3-25,l3-38} and about 0.6 below the world average ($21.07\pm0.11$)\cite{PDG2002}.
The difference in mean multiplicity between the cases of including or not the \PKzS\ and \PgL\ decay
products is consistent with our measurement of the \PKz\ and \PgL\ production rates \cite{l3-118}
and with the world average  \cite{PDG2002}.
All the moments, with the exception of the dispersion, $D$, show significant flavor dependence.
However, the flavor dependence of $F_2$ is quite small.
$F_2$ is also quite insensitive to the inclusion or not of \PKzS\ and \PgL\ decay products.
The difference between the mean charged-particle multiplicity
of the \Pqb-quark sample and that of the udsc-quark sample is
$2.58  \pm 0.03  \pm 0.08  $ when \PKzS\ and \PgL\ decay products are included and
$2.43  \pm 0.03  \pm 0.08  $ otherwise.
 
\renewcommand{\floatpagefraction}{.80}
\begin{table}[hbt]
\begin{center}
\begin{tabular}{|l||c@{$\pm$}c@{$\pm$}c|c@{$\pm$}c@{$\pm$}c|}\hline
\multicolumn{1}{|l||}{{\bf All events}}
   & \multicolumn{3}{c|}{without \PKzS\ and \PgL\ decay} &
   \multicolumn{3}{c|}{with \PKzS\ and \PgL\ decay} \\
\hline
$\langle  \Nch\rangle $                                  & 18.63 &  0.01  &  0.11  & 20.46 &  0.01  &  0.11 \\
$\langle\Nch^2\rangle $                                  & 381.7 &  0.3   &  4.4   & 457.7 &  0.3   &  4.9  \\
%$\langle\Nch^3\rangle $                                  & 8524  &  10    &  154   & 11108 &  12    &  181  \\
$\langle\Nch^3\rangle \cdot10^{-2}$                      & 85.2  &  0.1   &  1.5   & 111.1 &  0.1   &  1.8  \\
%$\langle\Nch^4\rangle $                                  &$205.9\cdot10^3$&$0.4\cdot10^3$&$5.1\cdot10^3$&
%                                                          $290.6\cdot10^3$&$0.5\cdot10^3$&$6.5\cdot10^3$  \\
$\langle\Nch^4\rangle \cdot10^{-3}$                      &205.9  &  0.4   &  5.1   & 290.6 &  0.5   & 6.5  \\
%$\langle\Nch^4\rangle $                                  &205918 &  362   &  5137  & 290551&  475   & 6529  \\
$D=\sqrt{\langle(\Nch-\langle \Nch\rangle )^2\rangle }$  & 5.888 &  0.005 &  0.051 & 6.244 &  0.005 &  0.051 \\
$S=\langle(\Nch-\langle  \Nch\rangle )^3\rangle /D^3$    & 0.596 &  0.004 &  0.010 & 0.600 &  0.004 &  0.010 \\
$K=\langle(\Nch-\langle  \Nch\rangle )^4\rangle /D^4 -3$ & 0.51  &  0.01  &  0.04  & 0.49  &  0.01  &  0.03  \\
$\langle \Nch\rangle/D$                                  & 3.164 &  0.002 &  0.016 & 3.277 &  0.002 &  0.016 \\
$F_2=\langle \Nch(\Nch-1)\rangle/\langle \Nch\rangle^2$ & 1.0461 &  0.0002 & 0.0040 & 1.0441& 0.0001&  0.0034 \\
\hline
\hline
\multicolumn{1}{|l||}{{\bf udsc-quark events}}
   & \multicolumn{3}{c|}{\ } & \multicolumn{3}{c|}{\ } \\
\hline
$\langle  \Nch\rangle $                                  & 18.07 &  0.01  &  0.21  & 19.88 &  0.01  &  0.21 \\
$\langle\Nch^2\rangle $                                  & 340.0 &  0.3   &  8.4   & 432.4 &  0.4   &  9.2  \\
%$\langle\Nch^3\rangle $                                  & 7826  &  11    &  273   & 10219 &  14    &  326  \\
$\langle\Nch^3\rangle  \cdot10^{-2}$                     & 78.3  &  0.1   &  2.7   & 102.2 &  0.1   &  3.3 \\
$\langle\Nch^4\rangle \cdot10^{-3}$                      &184.4  &  0.4   &  8.6   & 260.7 &  0.5   & 11.1  \\
%$\langle\Nch^4\rangle $                                  &184370 &  390   &  8593  & 260701&  531   & 11060 \\
$D=\sqrt{\langle(\Nch-\langle \Nch\rangle )^2\rangle }$  & 5.769 &  0.007 &  0.071 & 6.111 &  0.007 &  0.071 \\
$S=\langle(\Nch-\langle  \Nch\rangle )^3\rangle /D^3$    & 0.613 &  0.005 &  0.014 & 0.617 &  0.005 &  0.012 \\
$K=\langle(\Nch-\langle  \Nch\rangle )^4\rangle /D^4 -3$ & 0.54  &  0.02  &  0.06  & 0.53  &  0.02  &  0.05  \\
$\langle \Nch\rangle/D$                                  & 3.133 &  0.003 &  0.020 & 3.252 &  0.003 &  0.020 \\
$F_2=\langle \Nch(\Nch-1)\rangle/\langle \Nch\rangle^2$ & 1.0464 &  0.0002 & 0.0045 & 1.0441& 0.0002&  0.0038 \\
 
\hline
\hline
\multicolumn{1}{|l||}{{\bf \boldmath{$\Pqb$}-quark events}}
   & \multicolumn{3}{c|}{\ } & \multicolumn{3}{c|}{\ } \\
\hline
$\langle  \Nch\rangle $                                  & 20.51 &  0.02  &  0.19   & 22.45 &  0.03  &  0.19 \\
$\langle\Nch^2\rangle $                                  & 453.9 &  1.1   &  1.8    & 542.0 &  1.2   &  3.0  \\
%$\langle\Nch^3\rangle $                                  &10786  &  40    &   71    & 14006 &  48    &  111  \\
$\langle\Nch^3\rangle \cdot10^{-2}$                      &107.9  &  0.4   &   0.7   & 140.1 &  0.5   &  1.1 \\
$\langle\Nch^4\rangle \cdot10^{-3}$                      &273.9  &  1.5   &  1.9    & 385.8 &  1.9   & 1.7  \\
%$\langle\Nch^4\rangle $                                  &273891 &  1457  &  1866   & 385758& 1883   &  1725 \\
$D=\sqrt{\langle(\Nch-\langle \Nch\rangle )^2\rangle }$  & 5.78  &  0.01  &  0.07   & 6.16  &  0.01  &  0.07 \\
$S=\langle(\Nch-\langle  \Nch\rangle )^3\rangle /D^3$    & 0.574 &  0.017 &   0.008 & 0.573 &  0.017 & 0.007 \\
$K=\langle(\Nch-\langle  \Nch\rangle )^4\rangle /D^4 -3$ & 0.43  &  0.04  &   0.04  & 0.42  &  0.04  & 0.03 \\
$\langle \Nch\rangle/D$                                  & 3.551 &  0.006 &  0.055  & 3.645 &  0.005 & 0.049 \\
$F_2=\langle \Nch(\Nch-1)\rangle/\langle \Nch\rangle^2$ & 1.0305 &  0.0003 &0.0027  & 1.0307& 0.0002 & 0.0023 \\
\hline\end{tabular}\end{center}
\caption{Moments of the charged-particle multiplicity distribution for all, udsc-, and \Pqb-quark events.
The first uncertainty is statistical, the second systematic.
\label{tab:t1t3}
}
\end{table}
 
\subsection*{\boldmath{$H_q$}}
 
The \Hq\ are calculated from the unfolded charged-particle multiplicity distributions.
Since the $H_q$ are sensitive to  low statistics       at very high multiplicities,
we truncate the multiplicity distribution.
The \Hq\ thus obtained are biased estimators of the \Hq\ of the untruncated distribution.
This bias increases with stronger truncation, while the statistical uncertainty decreases,
which allows a more significant comparison with models.
It was  suggested~\cite{trunc} that
even without this truncation, the $H_q$ may be biased since a natural truncation occurs as a
consequence of the finiteness of the sample.
The truncation can induce oscillations or increase their size \cite{trunc}.
The truncation also introduces correlations between the \Hq, although
these are small for low $q$\cite{trunc,mangeol,opal-212}.
We choose the point of truncation such that multiplicities with relative error on $P(n)$ greater
than 50\% are rejected.
This corresponds, for all multiplicity distributions studied,
to about 0.005\% of events.
For all three samples (full, udsc, and b) the truncation is at
53 if  \PKzS\ and \PgL\ decay products are included in the multiplicity
and at 49 when they are not.
The \Hq\ presented here are calculated from distributions not including these decay products.
However, the \Hq\ are insensitive to their inclusion \cite{mangeol}.
 
\begin{figure}[htbp]
  \begin{center}
    \includegraphics[width=\figwidth]{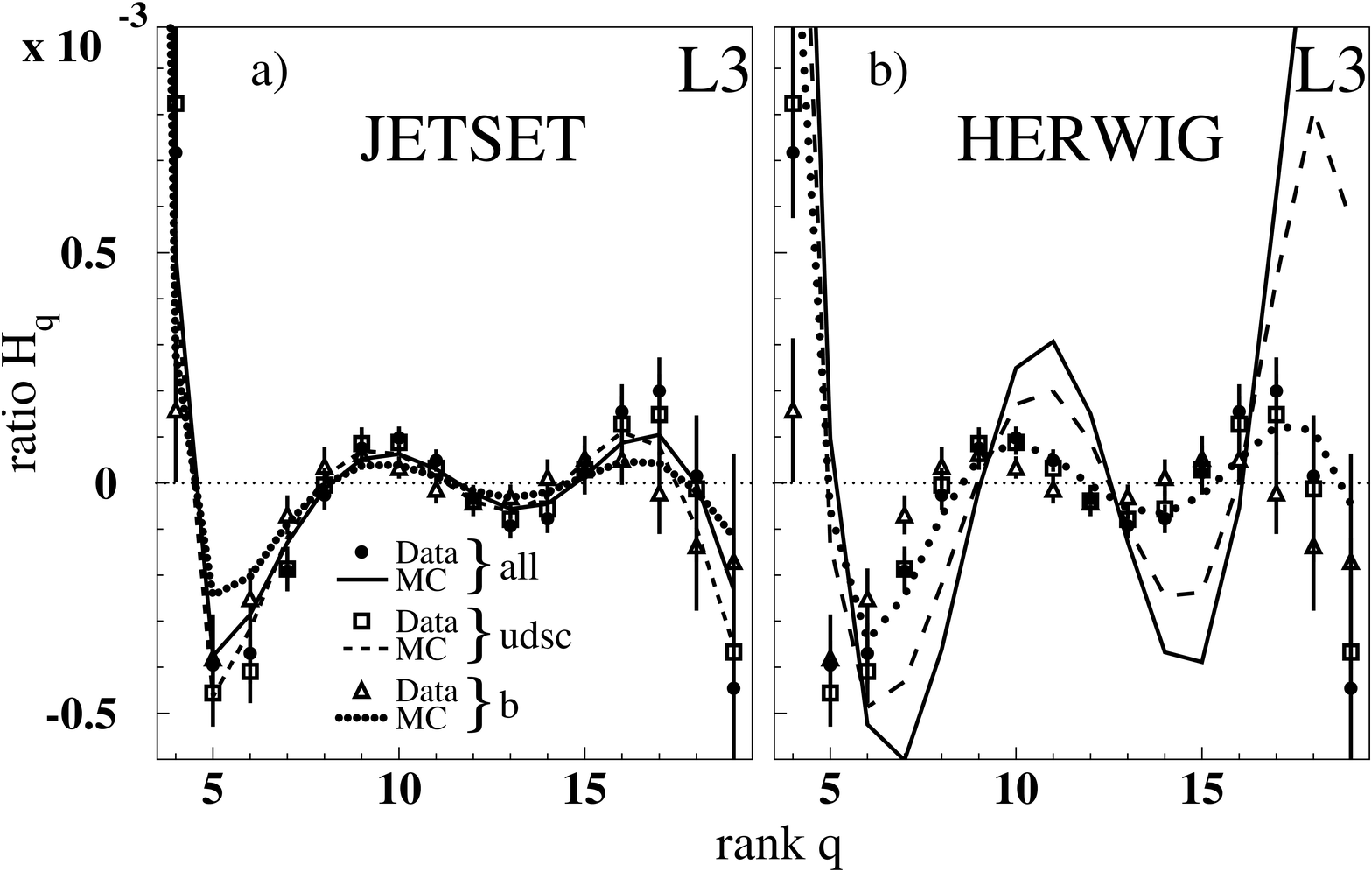}
    \includegraphics[width=\figwidth]{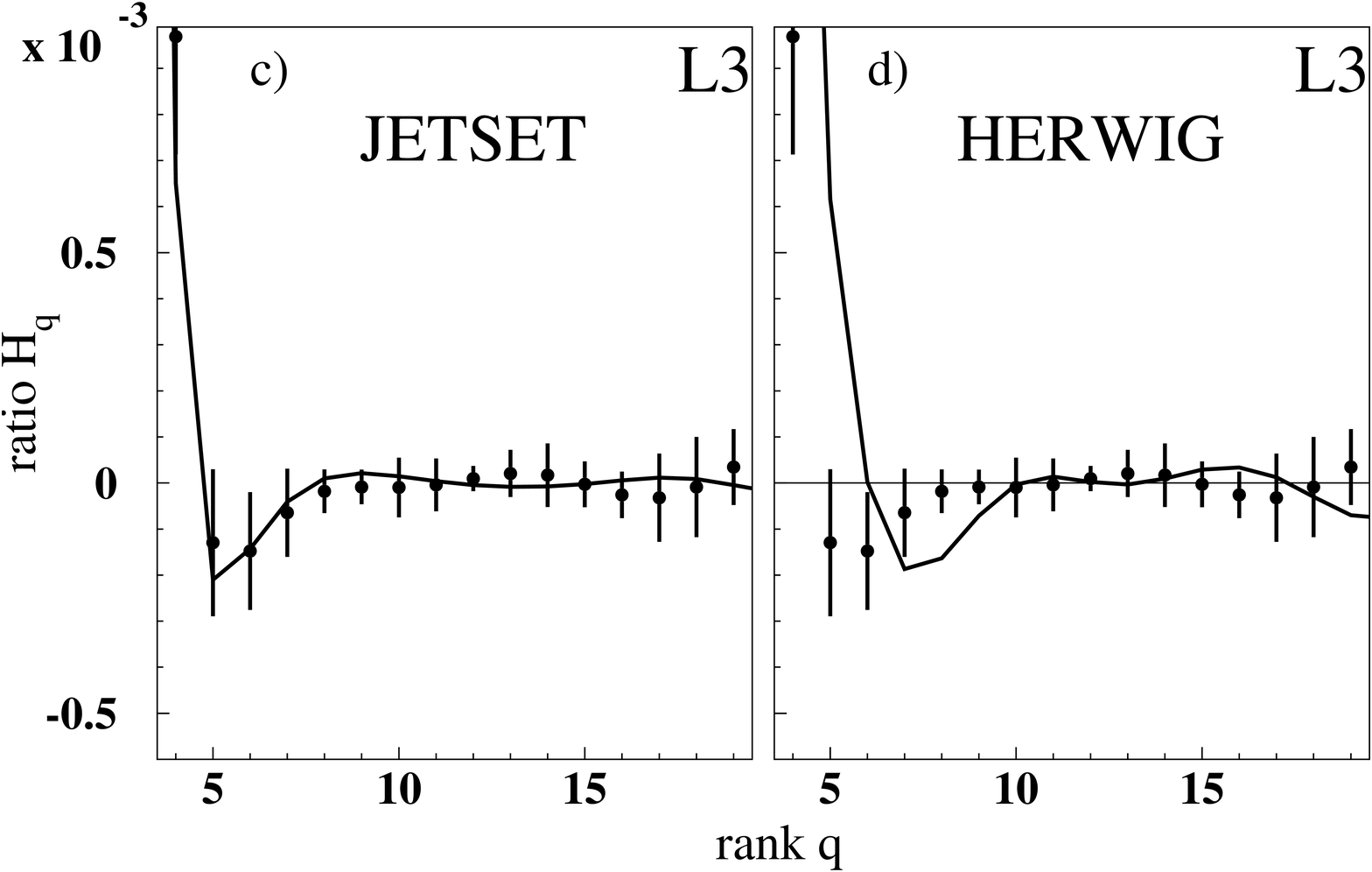}
  \end{center}
\icaption{
    The \Hq\ of the truncated (a,b) and non-truncated (c,d) charged-particle multiplicity distribution
    compared to the expectations of (a,c) \JETSET\ and (b,d) \HERWIG.
  The error bars include both statistical and systematic uncertainties.
  \label{fig:hqtrunc}
  \label{fig:hq}
}
\end{figure}

The \Hq\ of the truncated charged-particle multiplicity distribution from all,
udsc- and \Pqb-quark events, shown in Figure~\ref{fig:hqtrunc}, have a first
negative minimum at $q=5$ and quasi-oscillations for larger~$q$.
They are very similar for the three samples, with only slight differences
for the \Pqb-quark sample.
Similar behavior is seen for \JETSET\ (Figure~\ref{fig:hqtrunc}c).
Oscillations are also observed for \HERWIG\ (Figure~\ref{fig:hqtrunc}d),
but they do not agree with those seen in the data.
For both data and the Monte Carlo models,
truncation at a lower value increases the depth of the first minimum
and the amplitudes of the oscillations,
while truncation at a larger value has the opposite effect.
 
We note that our \Hq, based on an order of magnitude greater statistics,
agree with the \Hq\ of \SLD\ if we truncate
at a value equal to the maximum multiplicity they observed\cite{SLD}.

No truncation, other than that due to the finiteness of the sample, reduces the amplitudes
of the oscillations to statistical insignificance, but the minimum at $q=5$ remains, as is shown in
Figure~\ref{fig:hq}.  Again, \JETSET\ agrees well with the data, while \HERWIG\ does not.
 
\begin{figure}[htbp]
  \begin{center}
    \includegraphics[width=.9\figwidth]{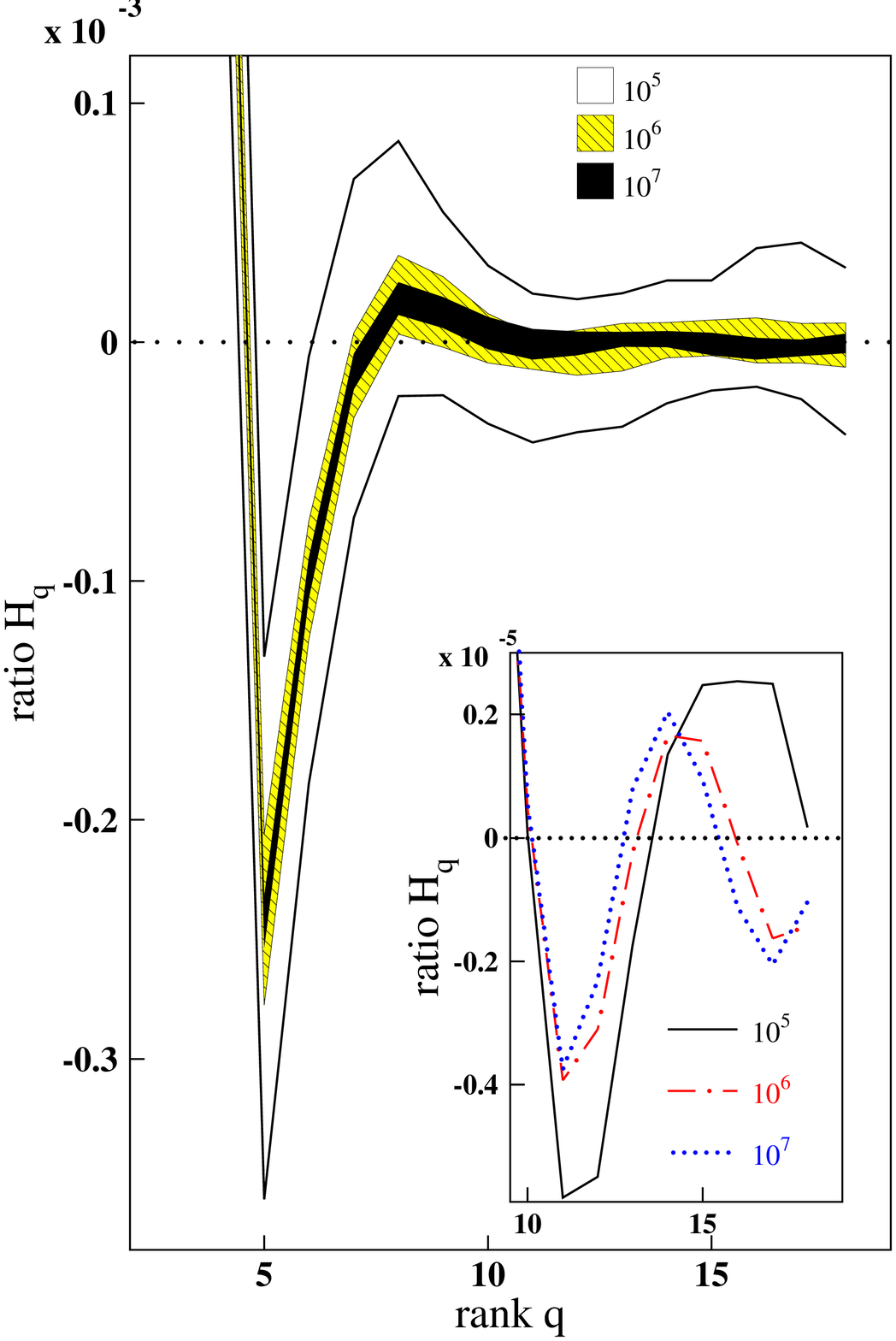}
  \end{center}
\icaption{
    1-standard deviation bands of  expectated  \Hq\ of the non-truncated charged-particle multiplicity
distribution from \PYTHIA\
for sample sizes of $10^5$, $10^6$ and $10^7$.
    The insert shows the mean \Hq\ of 100 samples of $10^5$, $10^6$ and $10^7$ \PYTHIA\ events.
  \label{fig:hqpythband}
}
\end{figure}

To investigate the effect of sample size on the \Hq,
100 samples of \PYTHIA\cite{PYTHIA61}  Monte Carlo events, were generated
for sample sizes of $10^5$, $10^6$ and $10^7$ events, and their \Hq\ determined.
Their $\pm1$ standard deviation bands are shown in Figure~\ref{fig:hqpythband}.
In the insert of Figure~\ref{fig:hqpythband} the mean of the values is shown.
For large $q$ the values of the $H_q$ depend on the sample size.
However, for small $q$ the values of the $H_q$ are stable.
In particular, $H_5$ (the first minimum) changes little with the sample size,
giving us confidence that the measured $H_5$ is robust.
Figure \ref{fig:hqpythband} suggests that at least $10^7$ events,  an order of magnitude beyond the
statistics of the present experiment, would be needed to establish the maximum at $q=8$.
 
%We conclude that the \Hq\ have a negative minimum at $q=5$ in qualitative agreement
%with MLLA and NNLLA and in quantitative agreement with \JETSET.
%However, the oscillatory behavior predicted in NNLLA cannot be confirmed.
%Previous observations of these oscillations are most likely a consequence of truncation resulting from
%limited statistics.
 
\section*{Conclusions}
The charged particle
multiplicity distribution of hadronic \PZ\ decay and its moments are  measured for light-quark and for
b-quark, as well as for all flavor events.
The \Hq\ moments of truncated multiplicity distributions, which have smaller statistical uncertainties than
those of the full distributions, are plotted {\it versus\/} the rank $q$.  A negative minimum is observed at
$q=5$ followed by quasi-oscillations about zero, which is qualitatively similar to the behavior expected in
NNLLA for the \Hq\ moments of the full multiplicity distribution.  Since Monte Carlo studies show that these
oscillations are magnified, or even created, by truncation of the multiplicity distribution, the \Hq\ are
also measured for the untruncated multiplicity distribution.  In this case the minimum at $q=5$, expected in
both MLLA and NNLLA, is confirmed.  But the oscillations at higher values of $q$, which are expected only in
NNLLA, cannot be confirmed.
Previous observations of these oscillations are most likely a consequence of truncation resulting from
limited statistics.
 
%We conclude that the \Hq\ have a negative minimum at $q=5$ in qualitative agreement
%with MLLA and NNLLA and in quantitative agreement with \JETSET.
%However, the oscillatory behavior predicted in NNLLA cannot be confirmed.
%Previous observations of these oscillations are most likely a consequence of truncation resulting from
%limited statistics.

%%%%%%%%%%%%%%%%%%%%%%%%%%%%%%%%%%%%%%%%%%%%%%%%%%%%%%%%%%%%%%%%%%%%%%%%%%%%%%%
% Acknowledgements
%%%%%%%%%%%%%%%%%%%%%%%%%%%%%%%%%%%%%%%%%%%%%%%%%%%%%%%%%%%%%%%%%%%%%%%%%%%%%%%
%
\section*{Acknowledgments}
We have benefited from discussions with I.~Dremin, W.~Ochs, A.~Giovannini and R.~Ugoccioni.

%
%%%%%%%%%%%%%%%%%%%%%%%%%%%%%%%%%%%%%%%%%%%%%%%%%%%%%%%%%%%%%%%%%%%%%%%%%%%%%%%
% Bibliography
%%%%%%%%%%%%%%%%%%%%%%%%%%%%%%%%%%%%%%%%%%%%%%%%%%%%%%%%%%%%%%%%%%%%%%%%%%%%%%
%

%%%%%%%%%%%%%%%%%%%%%%%%%%%%%%%%%%%%%%%%%%%%%%%%%%%%%%%%%%%%%%%%%%%%%%%%%%%%%%%
% The author list
%%%%%%%%%%%%%%%%%%%%%%%%%%%%%%%%%%%%%%%%%%%%%%%%%%%%%%%%%%%%%%%%%%%%%%%%%%%%%%%
%
%\section*{Author List}
%%Only after the paper has been approved at a 5 o'clock meeting should
%you get the number of the author list from Bob Clare and include it
%after the references.
\newpage
\typeout{   }     
\typeout{Using author list for paper 243 -- ? }
\typeout{$Modified: Jul 31 2001 by smele $}
\typeout{!!!!  This should only be used with document option a4p!!!!}
\typeout{   }
%
%
%
%  L A T E X  version!!
%
%
% Make sure that the Lep package has been used!
%\input{Lep.sty}%
%
%\ifx\LepCalled\undefined%
%\typeout{     }%
%\typeout{!!!!!!!!!!!!!!!!!!!!!!!!!!!!!!!!!!!!!!!!!!!!!!!!!!!!!!!!!!!}%
%\typeout{Yikes.  You haven't used the Lep package!}%
%\typeout{Please put \protect\usepackage\protect{Lep\protect} in your preamble,
%         followed by}%
%\typeout{\protect\Lep\protect{1\protect} or \protect\Lep\protect{2\protect}}%
%\typeout{     }%
%\typeout{For now you will get a Lep phase 2 authorlist (may not be right!).}%
%\typeout{!!!!!!!!!!!!!!!!!!!!!!!!!!!!!!!!!!!!!!!!!!!!!!!!!!!!!!!!!!!}%
%\typeout{     }%
%\Lep{2}\fi%

\newcount\tutecount  \tutecount=0
\def\tutenum#1{\global\advance\tutecount by 1 \xdef#1{\the\tutecount}}
\def\tute#1{$^{#1}$}
\tutenum\aachen            % 1
\tutenum\nikhef            % 2
\tutenum\mich              % 3
\tutenum\lapp              % 4
\tutenum\basel             % 5
\tutenum\lsu               % 6
\tutenum\beijing           % 7
\tutenum\berlin            % 8
\tutenum\bologna           % 9 
\tutenum\tata              % 10
\tutenum\ne                % 11
\tutenum\bucharest         % 12
\tutenum\budapest          % 13
\tutenum\mit               % 14 
\tutenum\panjab            % 15
\tutenum\debrecen          % 16
\tutenum\florence          % 17
\tutenum\cern              % 18
\tutenum\wl                % 19
\tutenum\geneva            % 20
\tutenum\hefei             % 21
\tutenum\lausanne          % 22
\tutenum\lyon              % 23
\tutenum\madrid            % 24
\tutenum\florida           % 25
\tutenum\milan             % 26
\tutenum\moscow            % 27
\tutenum\naples            % 29
\tutenum\cyprus            % 30
\tutenum\nymegen           % 31
\tutenum\caltech           % 32
\tutenum\perugia           % 33
\tutenum\peters            % 34
\tutenum\cmu               % 35
\tutenum\potenza           % 36
\tutenum\prince            % 37
\tutenum\riverside         % 38
\tutenum\rome              % 39
\tutenum\salerno           % 40
\tutenum\ucsd              % 41
\tutenum\sofia             % 42
\tutenum\korea             % 43
\tutenum\utrecht           % 44
\tutenum\purdue            % 45
\tutenum\psinst            % 46
\tutenum\zeuthen           % 47
\tutenum\eth               % 48
\tutenum\hamburg           % 49
\tutenum\taiwan            % 50
\tutenum\tsinghua          % 51

{
\parskip=0pt
\noindent
{\bf The L3 Collaboration:}
\ifx\selectfont\undefined%  old style font selection
 \baselineskip=10.8pt
 \baselineskip\baselinestretch\baselineskip
 \normalbaselineskip\baselineskip
 \ixpt
\else%                      new style font selection
 \fontsize{9}{10.8pt}\selectfont
\fi
\medskip
\tolerance=10000
\hbadness=5000
\raggedright
\hsize=162truemm\hoffset=0mm
\def\r{\rlap,}
\noindent

P.Achard\r\tute\geneva\ 
O.Adriani\r\tute{\florence}\ 
M.Aguilar-Benitez\r\tute\madrid\ 
J.Alcaraz\r\tute{\madrid,\cern}\ 
G.Alemanni\r\tute\lausanne\
J.Allaby\r\tute\cern\
A.Aloisio\r\tute\naples\ 
M.G.Alviggi\r\tute\naples\
H.Anderhub\r\tute\eth\ 
V.P.Andreev\r\tute{\lsu,\peters}\
% ...
\newpage
%\rule{\textwidth}{0.4pt}
\begin{list}{A}{\itemsep=0pt plus 0pt minus 0pt\parsep=0pt plus 0pt minus 0pt
                \topsep=0pt plus 0pt minus 0pt}
\item[\aachen]
 I. Physikalisches Institut, RWTH, D-52056 Aachen, FRG$^{\S}$\\
 III. Physikalisches Institut, RWTH, D-52056 Aachen, FRG$^{\S}$
\item[\nikhef] National Institute for High Energy Physics, NIKHEF, 
     and University of Amsterdam, NL-1009 DB Amsterdam, The Netherlands
\item[\mich] University of Michigan, Ann Arbor, MI 48109, USA
\item[\lapp] Laboratoire d'Annecy-le-Vieux de Physique des Particules, 
     LAPP,IN2P3-CNRS, BP 110, F-74941 Annecy-le-Vieux CEDEX, France
\item[\basel] Institute of Physics, University of Basel, CH-4056 Basel,
     Switzerland
\item[\lsu] Louisiana State University, Baton Rouge, LA 70803, USA
\item[\beijing] Institute of High Energy Physics, IHEP, 
  100039 Beijing, China$^{\triangle}$ 
\item[\berlin] Humboldt University, D-10099 Berlin, FRG$^{\S}$
\item[\bologna] University of Bologna and INFN-Sezione di Bologna, 
     I-40126 Bologna, Italy
\item[\tata] Tata Institute of Fundamental Research, Mumbai (Bombay) 400 005, India
\item[\ne] Northeastern University, Boston, MA 02115, USA
\item[\bucharest] Institute of Atomic Physics and University of Bucharest,
     R-76900 Bucharest, Romania
\item[\budapest] Central Research Institute for Physics of the 
     Hungarian Academy of Sciences, H-1525 Budapest 114, Hungary$^{\ddag}$
\item[\mit] Massachusetts Institute of Technology, Cambridge, MA 02139, USA
\item[\panjab] Panjab University, Chandigarh 160 014, India.
\item[\debrecen] KLTE-ATOMKI, H-4010 Debrecen, Hungary$^\P$
\item[\florence] INFN Sezione di Firenze and University of Florence, 
     I-50125 Florence, Italy
\item[\cern] European Laboratory for Particle Physics, CERN, 
     CH-1211 Geneva 23, Switzerland
\item[\wl] World Laboratory, FBLJA  Project, CH-1211 Geneva 23, Switzerland
\item[\geneva] University of Geneva, CH-1211 Geneva 4, Switzerland
\item[\hefei] Chinese University of Science and Technology, USTC,
      Hefei, Anhui 230 029, China$^{\triangle}$
\item[\lausanne] University of Lausanne, CH-1015 Lausanne, Switzerland
\item[\lyon] Institut de Physique Nucl\'eaire de Lyon, 
     IN2P3-CNRS,Universit\'e Claude Bernard, 
     F-69622 Villeurbanne, France
\item[\madrid] Centro de Investigaciones Energ{\'e}ticas, 
     Medioambientales y Tecnolog{\'\i}cas, CIEMAT, E-28040 Madrid,
     Spain${\flat}$ 
\item[\florida] Florida Institute of Technology, Melbourne, FL 32901, USA
\item[\milan] INFN-Sezione di Milano, I-20133 Milan, Italy
\item[\moscow] Institute of Theoretical and Experimental Physics, ITEP, 
     Moscow, Russia
\item[\naples] INFN-Sezione di Napoli and University of Naples, 
     I-80125 Naples, Italy
\item[\cyprus] Department of Physics, University of Cyprus,
     Nicosia, Cyprus
\item[\nymegen] University of Nijmegen and NIKHEF, 
     NL-6525 ED Nijmegen, The Netherlands
\item[\caltech] California Institute of Technology, Pasadena, CA 91125, USA
\item[\perugia] INFN-Sezione di Perugia and Universit\`a Degli 
     Studi di Perugia, I-06100 Perugia, Italy   
\item[\peters] Nuclear Physics Institute, St. Petersburg, Russia
\item[\cmu] Carnegie Mellon University, Pittsburgh, PA 15213, USA
\item[\potenza] INFN-Sezione di Napoli and University of Potenza, 
     I-85100 Potenza, Italy
\item[\prince] Princeton University, Princeton, NJ 08544, USA
\item[\riverside] University of Californa, Riverside, CA 92521, USA
\item[\rome] INFN-Sezione di Roma and University of Rome, ``La Sapienza",
     I-00185 Rome, Italy
\item[\salerno] University and INFN, Salerno, I-84100 Salerno, Italy
\item[\ucsd] University of California, San Diego, CA 92093, USA
\item[\sofia] Bulgarian Academy of Sciences, Central Lab.~of 
     Mechatronics and Instrumentation, BU-1113 Sofia, Bulgaria
\item[\korea]  The Center for High Energy Physics, 
     Kyungpook National University, 702-701 Taegu, Republic of Korea
\item[\utrecht] Utrecht University and NIKHEF, NL-3584 CB Utrecht, 
     The Netherlands
\item[\purdue] Purdue University, West Lafayette, IN 47907, USA
\item[\psinst] Paul Scherrer Institut, PSI, CH-5232 Villigen, Switzerland
\item[\zeuthen] DESY, D-15738 Zeuthen, 
     FRG
\item[\eth] Eidgen\"ossische Technische Hochschule, ETH Z\"urich,
     CH-8093 Z\"urich, Switzerland
\item[\hamburg] University of Hamburg, D-22761 Hamburg, FRG
\item[\taiwan] National Central University, Chung-Li, Taiwan, China
\item[\tsinghua] Department of Physics, National Tsing Hua University,
      Taiwan, China
\item[\S]  Supported by the German Bundesministerium 
        f\"ur Bildung, Wissenschaft, Forschung und Technologie
\item[\ddag] Supported by the Hungarian OTKA fund under contract
numbers T019181, F023259 and T024011.
\item[\P] Also supported by the Hungarian OTKA fund under contract
  number T026178.
\item[$\flat$] Supported also by the Comisi\'on Interministerial de Ciencia y 
        Tecnolog{\'\i}a.
\item[$\sharp$] Also supported by CONICET and Universidad Nacional de La Plata,
        CC 67, 1900 La Plata, Argentina.
\item[$\triangle$] Supported by the National Natural Science
  Foundation of China.
\end{list}
}
\vfill

%%% Local Variables: 
%%% mode: latex
%%% TeX-master: t
%%% End:

\newpage
%\listoffigures
 
%
%%%%%%%%%%%%%%%%%%%%%%%%%%%%%%%%%%%%%%%%%%%%%%%%%%%%%%%%%%%%%%%%%%%%%%%%%%%%%%
%
 
%
%%%%%%%%%%%%%%%%%%%%%%%%%%%%%%%%%%%%%%%%%%%%%%%%%%%%%%%%%%%%%%%%%%%%%%%%%%%%%%
%
 
%
%%%%%%%%%%%%%%%%%%%%%%%%%%%%%%%%%%%%%%%%%%%%%%%%%%%%%%%%%%%%%%%%%%%%%%%%%%%%%%
%
 
%
%%%%%%%%%%%%%%%%%%%%%%%%%%%%%%%%%%%%%%%%%%%%%%%%%%%%%%%%%%%%%%%%%%%%%%%%%%%%%%
%
 
%
%%%%%%%%%%%%%%%%%%%%%%%%%%%%%%%%%%%%%%%%%%%%%%%%%%%%%%%%%%%%%%%%%%%%%%%%%%%%%%
\end{document}